\documentclass[aps,pra,amsmath,amssymb,showpacs,letterpaper,twocolumn,superscriptaddress]{revtex4-1}%
\usepackage{graphicx}
\usepackage{amsmath}
\usepackage{graphicx}%
\usepackage{amsfonts}%
\usepackage{amssymb}
\usepackage{color}
\usepackage{setspace}
\usepackage{textcomp}
\usepackage{multirow}
\usepackage{hyperref}
\usepackage{booktabs}
\usepackage{pdfpages}
\makeatletter \AtBeginDocument{\let\LS@rot\@undefined} \makeatother
\usepackage{pgffor}

\begin{document}
	
\title{Multiscale Modeling of Ion Transport in Nanopores: Fitting Implicit-Water Radial Diffusion Profiles to Explicit-Water Molecular Dynamics} 

\author{M\'onika Valisk\'o}
\affiliation{Center for Natural Sciences, University of Pannonia, P.O. Box 158, H-8201 Veszpr\'em, Hungary}

\author{Salman Shabbir}
\affiliation{Center for Natural Sciences, University of Pannonia, P.O. Box 158, H-8201 Veszpr\'em, Hungary}
\affiliation{Department of Engineering, Reykjavik University, Menntavegur 1, 102 Reykjavík, Iceland}

\author{Eszter Moln\'{a}rn\'{e} Lakics}
\affiliation{Center for Natural Sciences, University of Pannonia, P.O. Box 158, H-8201 Veszpr\'em, Hungary}
	
\author{Zolt\'an Hat\'o}
\affiliation{Center for Natural Sciences, University of Pannonia, P.O. Box 158, H-8201 Veszpr\'em, Hungary}

\author{Dezs\H{o} Boda}\email[Author for correspondence:]{dezsoboda@gmail.com}
\affiliation{Center for Natural Sciences, University of Pannonia, P.O. Box 158, H-8201 Veszpr\'em, Hungary}

\date{\today}
	
% Keywords: activity coefficient, solvation, Grand Canonical Monte Carlo, Born energy
	
\begin{abstract}
We develop a multiscale approach for incorporating molecular-scale transport information into computationally efficient models of ion transport through nanopores.
A radially varying effective diffusion coefficient profile is fitted to radial conductivity profiles obtained from explicit-water molecular dynamics (MD) simulations. The fitting is performed within the NP+LEMC framework, which combines the Nernst–Planck equation with Local Equilibrium Monte Carlo to account for ion correlations beyond mean-field approximation. We apply the approach to NaCl, CaCl$_2$, and their mixtures in a negatively charged silica nanopore. The resulting diffusion coefficient profiles reproduce the radial current distributions of the MD simulations, including the strong suppression of ionic mobility near the pore wall that cannot be captured by a spatially constant diffusion coefficient inside the pore. NaCl and CaCl$_2$ exhibit qualitatively different transport behavior: the former is cation selective due to enhanced near-wall Na$^+$ conduction, whereas the latter shows weak anion selectivity because strongly bound Ca$^{2+}$ ions have strongly suppressed mobility near the surface. For NaCl--CaCl$_2$ mixtures, preferential Ca$^{2+}$ binding leads to nonlinear changes in ionic conductance. The approach establishes a bridge between experimentally relevant device behavior and computationally efficient reduced models, with explicit-water MD providing the molecular-scale information that can be incorporated into the effective transport coefficients.
%The resulting approach provides adjustable diffusion coefficient profiles that may reproduce not only the device-level behavior (informed, for example, by experiments), but also more accurate molecular-level behavior (informed by MD).
% \begin{center}
% \includegraphics*[width=0.4\textwidth]{figs_scaling/toc}\\
% TOC figure
% \end{center}
\end{abstract}
	
% \pacs{02.70.-c, 02.70.Uu, 05.10.Ln}
	
\maketitle
	
%#######################################################################
%#######################################################################
	
\clearpage
	
\section{Introduction}
To understand the behavior of confined aqueous electrolytes, particularly ionic transport in porous materials, a variety of methods are available. While experiments provide the ultimate benchmark for modeling studies, they generally yield results for integrated quantities, such as electric currents. Investigating local quantities, such as current densities and concentration profiles, or ionic correlations requires modeling the system of interest and studying it using statistical mechanical methods.

Depending on the resolution of the model and the accuracy of the methodology, different levels of description can be targeted within a multiscale modeling strategy.~\cite{steinhauser_multiscale_2008,hato_pccp_2017,matejczyk_jcp_2017,valisko_jcp_2019,boda_entropy_2020}
While computer simulations provide exact results for a given model, apart from finite-size effects and statistical uncertainties, theoretical approaches usually involve built-in approximations that limit the accuracy of the computed results.~\cite{allen_tildesley,Frenkel_Smit,sadus99}

In the case of nanoscale pores that facilitate the controlled transport of ions between two compartments separated by a membrane, the major divide in model resolution lies between all-atom models, including explicit water, and simplified models that use reduced representations of the electrolyte and membrane, including an implicit treatment of water (see Fig.\ 1 of Ref.~\cite{valisko_jcp_2019}).
While all-atom models are commonly studied using molecular dynamics (MD) simulations~\cite{rapaport}, a wider range of methodological choices is available for reduced models.
These models can be studied using Brownian dynamics (BD) simulations~\cite{van_gunsteren_mp_1982,berti_jctc_2014,Melnikov2023,farago_physicaA_2019} or continuum approaches based on the Nernst-Planck (NP) equation for the ionic current density:
\begin{equation}
	\mathbf{j}_{i}(\mathbf{r})= -\frac{1}{kT} D_{i}(\mathbf{r})c_{i}(\mathbf{r})\nabla \mu_{i}(\mathbf{r}) .
	\label{eq:NP}
\end{equation}
Here, $i$ denotes the ion species,  $k$ is the Boltzmann constant, $T$ is the temperature, $\mathbf{j}_{i}(\mathbf{r})$ is the particle current density, $D_{i}(\mathbf{r})$ is the diffusion coefficient profile, $c_{i}(\mathbf{r})$ is the concentration profile, and $\mu_{i}(\mathbf{r})$ is the electrochemical potential profile.

To solve this equation, a closure relation between $c_{i}(\mathbf{r})$ and $\mu_{i}(\mathbf{r})$ is required and must be supplied by a statistical mechanical method.
This may be the mean-field Poisson-Boltzmann (PB) theory, which, when coupled to the NP equation, yields the Poisson-Nernst-Planck (PNP) theory,~\cite{Nonner1998,Kurnikova1999,Corry2000,Noskov2004,Constantin2007,Burger2012,Chaudhry2014,Liu2015,matejczyk_jcp_2017} or a method capable of accounting for ionic correlations beyond the mean-field approximation of PB, with Monte Carlo (MC) simulations~\cite{Im2000,boda_jctc_2012} and density functional theory~\cite{Gillespie2002,gillespie_pre_2003} being the main examples.

The accuracy with which concentration profiles, representing the configurational degrees of freedom, are reproduced depends primarily on the underlying model.
Explicit-water and implicit-water models inherently yield different ion distributions because of their fundamentally different treatments of water molecules.
The model can be refined by introducing more detailed representations of surface groups~\cite{lakics_jcp_2026} or by including dielectric boundaries,~\cite{corry-bj-84-3594-2003,sawada_pre_2016,nonner_bj_2000,boda_bj_2007,boda_prl_2007,boda_jgp_2009,csanyi_bba_2012,boda_jcp_2013_solvation} but the implicit treatment of water imposes an intrinsic limit on microscopic accuracy.

If, for a given reduced model, we wish to fit its results to a ``more accurate'' reference, such as experimental data ~\cite{almers_jp_1984b,corry-bj-2001,gillespie_jpcb_2005,gillespie_bj_2008_energetics,gillespie_giri_bj_2009,boda_entropy_2020,gillespie_bj_2008_ca,fabian_jml_2022,lakics_jcp_2026} or a more detailed molecular model~\cite{hato_cmp_2016,hato_pccp_2017,valisko_jcp_2019}, the remaining option is to adjust the diffusion coefficient profile so as to reproduce the results provided by the reference method. 

The use of position-dependent diffusion coefficients has a long history in the description of diffusion through confined and inhomogeneous systems. Early theoretical studies showed that the diffusion coefficient in a reduced description need not be a material constant, but can vary with position as a result of confinement and spatially varying interactions.~\cite{levitt_bj_1978,levitt_ar_1986,zwanzig_jpc_1992,reguera_pre_2001,hummer_njp_2005,kalinay_jcp_2005}
This concept was subsequently adopted in reduced descriptions of transport through membranes and ion channels, including BD approaches, where spatially varying diffusion or friction coefficients were used to account for the heterogeneous molecular environment and confinement.~\cite{Allen1999,moy-bj-2000,Corry2000,corry-bj-2001,corry-bj-84-3594-2003,Noskov2004,allen_jgp_2004,roux_2005,Melnikov2023}
Importantly, when a sufficiently broad range of experimental data is available, effective diffusion coefficients can prove transferable across different state points, including electrolyte compositions and applied voltages.~\cite{gillespie_jpcb_2005,gillespie_bj_2008_energetics,gillespie_giri_bj_2009,boda_entropy_2020}

Consequently, the resulting effective diffusion coefficient should not be interpreted as the true microscopic diffusion coefficient.~\cite{levitt_bj_1978,levitt_ar_1986}
Rather, it is an adjustable effective parameter that incorporates, in an aggregate manner, the effects of physical features and processes absent from the reduced model.
These include the coarse-grained treatment of water, the membrane, and surface groups; confinement effects; momentum-exchange mechanisms absent from the NP description; polarization effects; and other molecular-scale contributions. 
Effects associated with larger-scale geometrical features, such as access resistance~\cite{aguilella-arzo05} and pore length~\cite{lakics_jcp_2026}, can also be incorporated into the diffusion coefficient. 

The first question that arises is what kind of data should be used to fit the $D_{i}(\mathbf{r})$ profile.
In general, our goal is to reproduce and interpret (sometimes predict) experimental data.
Experiments, however, typically do not provide access to local quantities such as those appearing in Eq.~\ref{eq:NP}.
Fitting only to integrated quantities therefore limits our ability to gain insight into the underlying molecular mechanisms.

The obvious solution is to fit the reduced model to local quantities obtained from a more detailed model, typically an all-atom model studied with MD simulations.
We should keep in mind, however, that comparison with MD is not a comparison with reality, but rather with another model.
Even if the MD simulations reproduce integrated experimental observables, we cannot be certain that every relevant aspect of the system is accurately represented by the all-atom model.
This is the price we have to pay, but the potential gain is substantial.

If the reduced model can be fitted to the all-atom model, we obtain a computationally efficient model that can be used to explore a wider parameter range than is practically accessible to MD, most notably at low concentrations.~\cite{corry-bj-2001,boda_jpcb_2000,boda_entropy_2020,madai_jcp_2017}
It also enables simulations over larger length and time scales at considerably lower computational cost.

In either case, fitting spatially varying diffusion coefficient profiles in a PNP approach, or, equivalently, friction coefficient profiles in BD through the Einstein relation, $\gamma_{i}(\mathbf{r})=kT/D_{i}(\mathbf{r})$, is a well-established multiscale strategy.~\cite{chung-bj-77-2517-1999}
One approach is to incorporate information about diffusion coefficients in confined geometries obtained from MD simulations directly into the reduced model, namely, into the diffusion coefficient profile inside the pore.~\cite{corry-bj-2001,liu_jpcb_2004,rybka_fpe_2016}
While calculation of local diffusion coefficients in MD is cumbersome \cite{yeh_jpcb_2004,liu_jpcb_2004,Siboulet_2013,rezlerova_pccp_2023,lisal_ms_2013,harris_jcp_2026,harris_jcp_2026}, it is not ensured that it agrees with effective diffusion coefficient used in the NP equation.

Therefore, the inverse route is more promising.  
In this approach, the diffusion coefficient profile inside the pore is determined by fitting it to quantities obtained either from experiments or from MD simulations.~\cite{chung-bj-77-2517-1999,corry-bj-2001,gillespie_jpcb_2005,gillespie_bj_2008_energetics,gillespie_giri_bj_2009,boda_entropy_2020,valisko_jcp_2019,hato_pccp_2017}

To the best of our knowledge, however, previous studies ~\cite{Allen1999,moy-bj-2000,Corry2000,corry-bj-2001,corry-bj-84-3594-2003,Noskov2004,allen_jgp_2004,roux_2005,Melnikov2023,gillespie_jpcb_2005,gillespie_bj_2008_energetics,gillespie_giri_bj_2009,boda_entropy_2020,valisko_jcp_2019,lakics_jcp_2026,fabian_jml_2022} have considered spatial variations of the diffusion coefficient only in the axial direction, $D_{i}(z)$. Typically, experimental bulk diffusion coefficients were used outside the channel, while reduced fractions of these values were assigned to different regions inside the pore, such as the selectivity filter and channel vestibules or, more generally, regions with different pore radii and charges.~\cite{bemporad_bj_2004}

In the present work, we take this multiscale approach one step further by fitting a radially varying diffusion coefficient profile inside the pore, $D_{i}^{\mathrm{P}}(r)$, to reproduce the $z$-component of the radial ionic current density profile, $j_{i}(r)$, obtained from all-atom MD simulations of a silica nanopore (or, rather, the radial conductivity profile, $\kappa_{i}(r)=j_{i}(r)/E^{\mathrm{P}}$, where $E^{\mathrm{P}}$ is the average electric field strength inside the pore).
Rather than calibrating the reduced model against a single spatially averaged transport quantity within the pore, our approach uses the full radially resolved current distribution as the fitting target.
This establishes a direct connection between the microscopic spatial structure of ion transport revealed by explicit-water MD simulations and the effective transport coefficients of an implicit-solvent NP description.

The importance of radial variation cannot be overstated in a nanopore, whose defining characteristic is the formation of an electrical double layer (EDL) near the charged pore surface.
The width of the EDL, commonly characterized by the Debye screening length relative to the pore radius, $R^{\mathrm{P}}$, together with the surface charge density, is a primary determinant of the pore's selectivity for cations over anions.~\cite{bazant_pre_2004,zangle_csr_2010,bocquet_chemsocrev_2010,sarkadi_jcp_2021,sarkadi_jml_2022,sarkadi_jml_2023,sarkadi_jpcb_2026}

A well-established conceptual picture is to divide the pore interior radially into a surface region near the pore wall, corresponding approximately to the EDL, and a bulk-like region in the pore interior, provided such a region exists.
The latter is absent when the EDLs overlap.
Accordingly, the total conductance of a nanopore is commonly decomposed into surface and volume contributions.~\cite{bikerman_1940,wiersema_jcis_1966,obrien_jcsf_1978,obrien_cjc_1981,dukhin_advcollsci_1993,lyklema_book_1995,lyklema_csa_1998}

Experiments have shown that the selectivity of a negatively charged nanopore changes markedly when a monovalent electrolyte, such as KCl, is replaced by electrolytes containing multivalent cations, such as CaCl$_{2}$ or CoSepCl$_{3}$.~\cite{he_jacs_2009}
While KCl exhibits classical EDL behavior, resulting in cation selectivity, CoSepCl$_{3}$, containing trivalent cations, leads to anion selectivity as a consequence of charge inversion.
The divalent CaCl$_{2}$ system exhibits intermediate behavior: in the presence of Ca$^{2+}$ ions, the negatively charged pore becomes essentially non-selective.

This behavior was investigated in our previous work for a silica nanopore containing a 1 M CaCl$_{2}$ solution.~\cite{shabbir_jcp_2026}
Based on our MD simulations of bulk CaCl$_{2}$ electrolytes,~\cite{salman_jml_2025} we identified the ECCR2 force field for Ca$^{2+}$ and Cl$^{-}$ ions,~\cite{martinek_jcp_2018} combined with the TIP4P/2005 water model,~\cite{abascal_jcp_2005} as the most appropriate for reproducing the transport behavior of these ions.

Using this electrolyte model in a negatively charged silica nanopore, we found that Ca$^{2+}$ ions bind strongly to the deprotonated silanol groups on the pore wall, effectively neutralizing its surface charge.
Overcharging and charge inversion are practically absent; they are not sufficiently pronounced to produce anion selectivity.
The radial cation concentration profile exhibits a pronounced peak near the pore wall, whereas the velocity profiles, which reflect ionic mobility, show the opposite trend, decreasing as the pore wall is approached ($r\to R^{\mathrm{P}}$).
The resulting particle current density profiles exhibit a similar decrease near the wall (Fig.\ 1B of Ref.~\onlinecite{shabbir_jcp_2026}).

In summary, pore conductance is dominated by the volume contribution, whereas the surface contribution is small because of the suppressed mobility of Ca$^{2+}$ ions near the pore wall.
Volume conductance, on the other hand, is bulk-like from the perspective of selectivity in the absence of charge inversion.

When this system is simulated using an implicit-water reduced model with a constant diffusion coefficient inside the pore, $D_{i}^{\mathrm{P}}$, the current density profile exhibits the opposite trend, increasing toward the pore wall. Since the radial dependence of $\nabla \mu_{i}(\mathbf{r})$ is negligible, the current density profile largely follows the concentration profile (see Eq.~\ref{eq:NP}). This behavior is opposite to that predicted by the explicit-water MD simulations. We expect the MD result to be more consistent with experimental observations and therefore use the MD profiles as reference data for fitting radially varying diffusion coefficient profiles, $D^{\mathrm{P}}_{i}(r)$.

While our earlier studies showed that averaged diffusion coefficients inside the pore can be fitted to reproduce experimental trends,~\cite{boda_entropy_2020,fabian_jml_2022,lakics_jcp_2026} the microscopic behavior of the resulting reduced model does not reproduce that observed in explicit-water MD simulations, suggesting that it may not accurately represent the underlying microscopic behavior of the system.
To obtain a $j_{i}(r)$ profile with the appropriate microscopic behavior, it is therefore necessary to fit a radially varying diffusion coefficient profile, $D_{i}^{\mathrm{P}}(r)$, that incorporates the suppressed mobility of ions near the pore wall.

We perform this analysis using NaCl and CaCl$_{2}$ solutions and their mixtures of varying composition inside negatively charged silica nanopores. 
The NaCl--CaCl$_{2}$ mole-fraction study is motivated by the anomalous mole fraction effect (AMFE) observed by Gillespie et al.~\cite{gillespie_bj_2008_nanopore} in a biconical PET nanopore, where the total conductance exhibited a minimum as a function of the Ca$^{2+}$ mole fraction.

AMFE was originally observed in calcium channels~\cite{almers_jp_1984a,almers_jp_1984b} and has since been extensively studied using various theoretical and simulation approaches.~\cite{nonner_bj_2000,boda_jpcb_2000,boda_jcp_2006,boda_prl_2007,gillespie_bj_2008_ca,boda_jgp_2009,malasics_bba_2010_trivalent,boda_jcp_2011_analyze,boda_jcp_2013_solvation,gillespie_bj_2008_energetics,gillespie_giri_bj_2009,boda_entropy_2020,fabian_jml_2022,lakics_jcp_2026} 
It is commonly associated with the preferential binding of Ca$^{2+}$ ions, which displace monovalent cations from the narrow channel while their own contribution to the current remains limited at low bulk concentrations. 
The observation of AMFE in the relatively wide nanopore studied by Gillespie et al.~\cite{gillespie_bj_2008_nanopore} was therefore particularly interesting. 
While preferential Ca$^{2+}$ binding was also invoked to explain this behavior, our recent work suggested that Cl$^{-}$ leakage may additionally play an important role.~\cite{lakics_jcp_2026}

Apart from analyzing the radial dependence of $\kappa_{i}(r)$ and $D_{i}^{\mathrm{P}}(r)$, one of the motivations of the present study is therefore to examine whether an AMFE also occurs in the present system.

\section{Models and methods}

\subsection{Explicit-water MD simulations}

In Ref.~\cite{salman_jml_2025}, we tested various force fields and found that the ECCR2 model of Martinek et al.~\cite{martinek_jcp_2018}, combined with the TIP4P/2005 water model~\cite{abascal_jcp_2005}, is suitable for reproducing the bulk conductance, diffusion coefficient, viscosity, and neutron scattering data of 1 M CaCl$_{2}$. 
In this model, the charges of the ions (and other charged particles) are scaled according to $q_{\text{scaled}}=q_{\text{original}}/\sqrt{\epsilon_{\infty}}=0.75\,q_{\text{original}}$ to approximate the effects of electronic polarization and charge transfer,~\cite{leontyev_jcp_2009,leontyev_pccp_2011} where $\epsilon_{\infty}$ is the high-frequency dielectric constant. 
The distance parameters of the Lennard-Jones potential, which characterize the effective ionic diameters, are also reduced ($d_{2+}=0.267$ nm and $d_{-}=0.41$ nm). A similar scaled-charge force field has been applied to NaCl with $d_{+}=0.2115$ nm.~\cite{kohagen_jpcb_2014,martinek_jcp_2018,oostenbrink_jcc_2004,dang_jcp_1995,kohagen_jpcb_2015} 
Scaled-charge models were extensively tested for bulk electrolytes~\cite{kann_jcp_2014,biriukov_pccp_2018,yue_mp_2019,kohagen_jpcb_2014,martinek_jcp_2018,zeron_jcp_2019,duboue_jcp_2020,predota_jml_2020} and under confinement.~\cite{vazdar_jpcb_2013,kohagen_jpcl_2014,melcrova_sr_2016,magarkar_jpcl_2017,biriukov_pccp_2018,biriukov_jpcc_2020,lebreton_jcp_2020}

\begin{figure}[t!]
	\begin{center}
		\includegraphics*[width=0.35\textwidth]{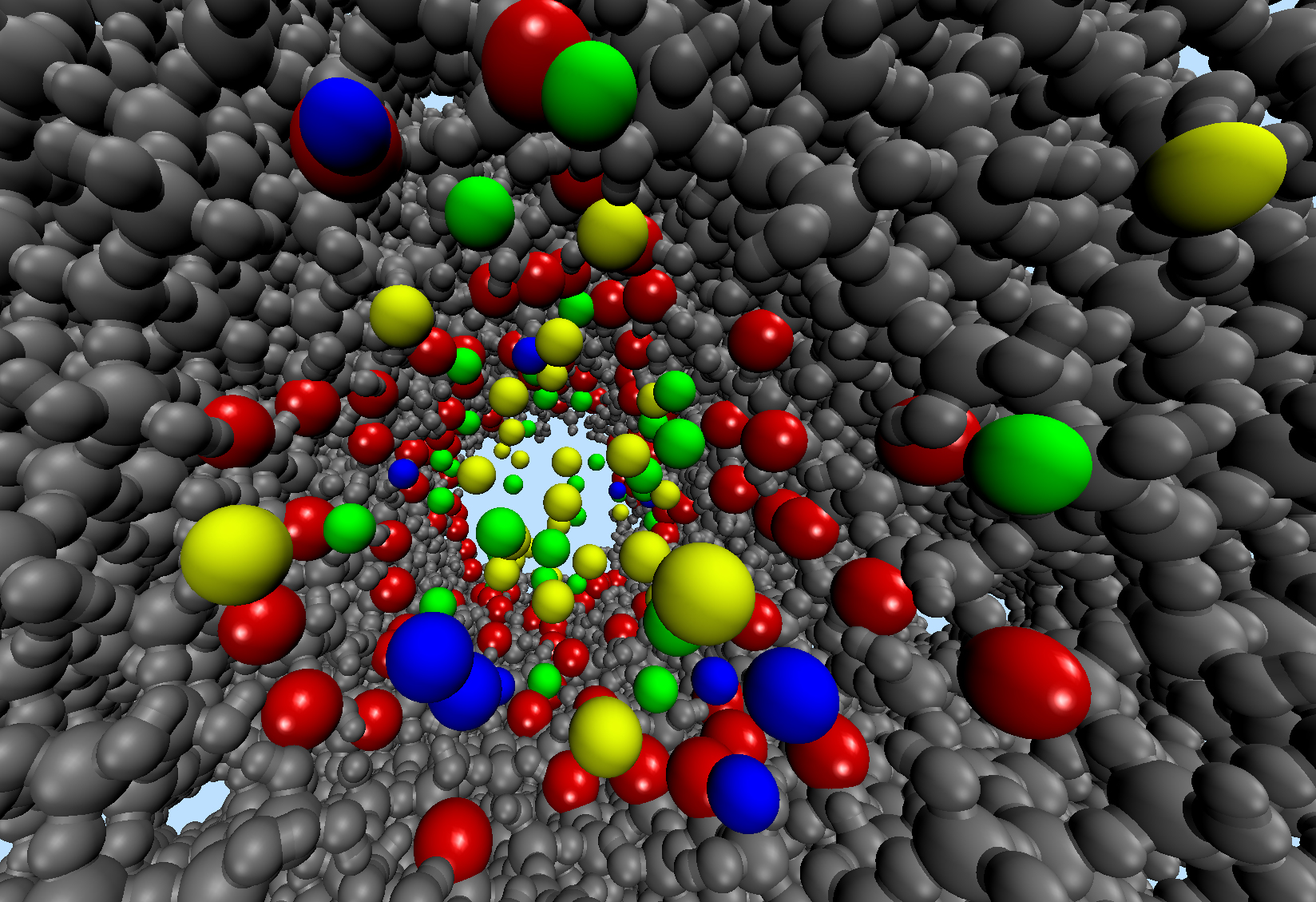}\\ \vspace{0.5cm}
		\includegraphics*[width=0.35\textwidth]{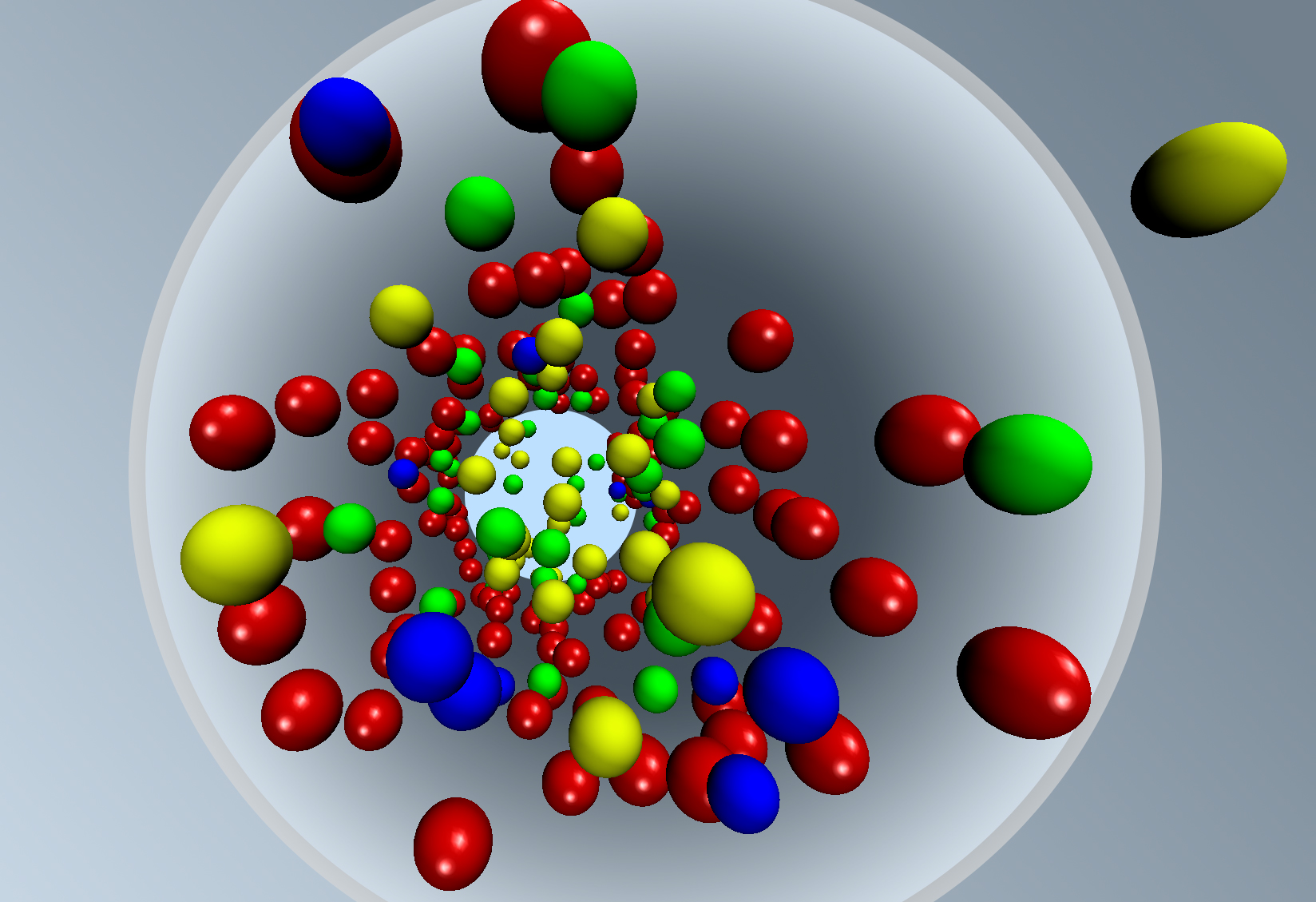}\\
	\end{center}
	\caption{Snapshots from (A) the MD simulation and (B) the LEMC simulation. Red, green, blue, and yellow spheres represent silanol oxygens, Ca$^{2+}$, Na$^{+}$, and Cl$^{-}$, respectively. The two simulation setups are otherwise closely matched; the LEMC model replaces explicit water molecules with an implicit solvent and represents the membrane by smooth hard walls rather than by an atomistic model.
	}
	\label{fig1}
\end{figure}

The silica pore was constructed using the Python package PoreMS.~\cite{PoreMS_Kraus_2021} 
The pore was carved out from a silica block and a subset of the dangling silanol groups was deprotonated. 
This procedure resulted in $104$ negatively charged silanol groups on the pore surface ($97$ isolated and $7$ geminal oxygens represented by red spheres in Fig.~\ref{fig1}A). 
The average radial distance of the oxygen atoms of these groups (O$_{\mathrm{S}}$) from the pore axis is $\approx 1.3$ nm. 
This arrangement corresponds to a surface group density of $\approx 1.54$/nm$^2$ and, using the charge of $-0.74e$ assigned to the O$_{\mathrm{S}}$ atoms by PoreMS, to a surface charge density of $\approx -1.14$ $e$/nm$^{2}$.

We further scale the charges of the O$_{\mathrm{S}}$ atoms to treat them consistently with the ECCR2 model of the ions, resulting in a charge of $-0.555e$ per O$_{\mathrm{S}}$ atom and a surface charge density of $\approx -0.85\,e$/nm$^{2}$. The number of surface groups remains unchanged for a given pore radius.
The results for this case are reported in the Supplementary Material (SM).

While our previous work \cite{shabbir_jcp_2026} reported results for pure NaCl and CaCl$_{2}$ solutions in the nanopore, here we also present results for mixtures of NaCl and CaCl$_{2}$, with the total salt concentration, $c_{\mathrm{tot}}=[\mathrm{NaCl}]+[\mathrm{CaCl}_{2}]$, kept constant at $\approx1$ M. 
The mole fraction of CaCl$_{2}$ is defined as $\eta=[\mathrm{CaCl}_{2}]/c_{\mathrm{tot}}$.

The MD simulations were performed using the GROMACS molecular simulation software suite, version 2023.2.~\cite{hess_jctc_2008} 
The dimensions of the rectangular simulation cell are $L\approx 7.1$ nm and $H\approx 18.4$ nm parallel and perpendicular to the membrane, respectively.
The radius and length of the pore was $R^{\mathrm{P}}\approx 1.3$ nm and $H^{\mathrm{P}}\approx 7.9$ nm, respectively; these are approximated values due to the ragged surfaces.

A homogeneous external electric field of strength $E=0.06606$ V/nm was applied along the $z$-axis; this value lies within the linear-response regime. Positions and velocities were recorded every $5$ ps for post-processing analysis during a $100$ ns production run. 
In our previous work, we demonstrated that this simulation length is sufficient to obtain converged results.~\cite{shabbir_jcp_2026}
Further details of the simulation setup and protocol can be found in our previous works.~\cite{salman_jml_2025,shabbir_jcp_2026}

In this study, we average over the azimuthal angle and along the pore length, $|z|<H^{\mathrm{P}}/2$, to obtain the radial profiles $\mathbf{j}_{i}(r)$ and $c_{i}(r)$ inside the pore. 
Since ion transport is driven along the pore axis, we report the $z$-component of the current density denoted concisely as $j_{i}(r)$.

After dividing the pore interior into volume elements in the radial direction, the velocity of ionic species $i$ in volume element $\alpha$ is computed as
\begin{equation}
	v_{i}^{\alpha}=\dfrac{1}{N_{i}^{\alpha}}\sum_{k=1}^{N_{i}^{\alpha}}\dfrac{\Delta z_{i,k}}{\Delta t},
	\label{eq:vialpha}
\end{equation}
where $N_{i}^{\alpha}\ne 0$ is the total number of occurrences of ions of species $i$ in subvolume $\alpha$ over the sampled configurations, and $\Delta z_{i,k}$ is the displacement of the corresponding ion in the $z$ direction during the time interval $\Delta t$.
The concentration in volume element $\alpha$ is obtained as
\begin{equation}
	c_{i}^{\alpha}=\dfrac{N_{i}^{\alpha}}{N_{t}V^{\alpha}},
	\label{eq:cialpha}
\end{equation}
where $N_{t}$ is the number of sampled configurations and $V^{\alpha}$ is the volume of subvolume $\alpha$.
As shown in the Appendix of our previous work,~\cite{shabbir_jcp_2026} the corresponding particle current density is
\begin{equation}
	j_{i}^{\alpha}= v_{i}^{\alpha}c_{i}^{\alpha} =\dfrac{1}{N_{t}V^{\alpha}}\sum_{k=1}^{N_{i}^{\alpha}}\dfrac{\Delta z_{i,k}}{\Delta t}.
	\label{eq:jialpha}
\end{equation}
The particle current, $J_{i}$, is obtained by integrating $j_{i}(r)$ over the cross section of the pore. 
The corresponding charge current is obtained as $I_{i}=q_{i}J_{i}$, where $q_{i}$ is the charge of ionic species $i$.

\subsection{Implicit-water NP+LEMC simulations}

In the reduced model, we aim to mimic the MD model as closely as possible while retaining only the essential degrees of freedom (see Fig.~\ref{fig1}).
These degrees of freedom are treated explicitly, whereas the effects of the remaining degrees of freedom are incorporated in an averaged manner, for example, through appropriate response functions.

The ions are modeled as charged hard spheres with point charges located at their centers. 
The pair interaction between ions of species $i$ and $j$ is given by
\begin{equation}
	u_{ij}(r) =
	\left\lbrace 
	\begin{array}{ll}
		\infty & \quad \mathrm{for} \quad r<R_{i}+R_{j} \\
		\dfrac{1}{4\pi\epsilon_{0}\epsilon} \dfrac{q_{i}q_{j}}{r} & \quad \mathrm{for} \quad r \geq R_{i}+R_{j}\\
	\end{array}
	\right. 
	\label{eq:uij},
\end{equation}
where $R_{i}$ and $R_{j}$ are the radii of ionic species $i$ and $j$, respectively, $r$ is the distance between the ion centers, $\epsilon_0$ is the vacuum permittivity, and $q_{i}$ and $q_{j}$ are the ionic charges. 
%The charge of species $i$ is given by $q_{i}=z_{i}e$, where $z_{i}$ is the ionic valence and $e$ is the elementary charge. 
The Pauling radii $0.095$, $0.099$, and $0.181$ were used for Na$^{+}$, Ca$^{2+}$, and Cl$^{-}$, respectively.

The dielectric constant, $\epsilon=78.15$, in Eq.~\ref{eq:uij} acts as a response function that accounts for the dielectric screening provided by water molecules.
Another response function is the diffusion coefficient profile, $D_{i}(z,r)$, which characterizes the effective friction experienced by the ions due to their environment.
While the diffusion coefficient is closely related to ionic mobility in a dilute bulk solution, in a concentrated and confined system such as that considered here, $D_{i}(z,r)$ has a broader interpretation.
It represents an effective transport coefficient that incorporates, in an averaged manner, the effects of degrees of freedom and physical mechanisms absent from the reduced model.

In this work, the diffusion coefficient profile is defined as  
\begin{equation}
	D_i (z,r) = \left\lbrace 
	\begin{array}{ll}
		D_i^{\mathrm{P}}(r) & \quad \mathrm{for} \quad |z|< H^{\mathrm{P}}/2 \\
		D_i^{\mathrm{B}}  & \quad  \mathrm{for} \quad |z|\geq H^{\mathrm{P}}/2\\
	\end{array}
	\right. 
	\label{eq:diffconst}
\end{equation}
where $H^{\mathrm{P}}$ is the length of the pore. 
Superscripts B and P refer to bulk and pore, respectively.
Experimental diffusion coefficients ($D_{i}^{\mathrm{B}}=1.334\times 10^{-9}$ m$^{2}$/s, $7.92\times 10^{-10}$ m$^{2}$/s, and $2.032\times 10^{-9}$ m$^{2}$/s for Na$^{+}$, Ca$^{2+}$, and Cl$^{-}$, respectively) are used in the bulk phases, whereas a radially varying diffusion coefficient, $D_{i}^{\mathrm{P}}(r)$, is used inside the pore.

The simulation cell is a cylinder of length $H=20$ nm and radius $R=3.5$ nm.
The membrane is represented by hard walls separated by a distance $H^{\mathrm{P}}=8$ nm, with overlap between the hard-sphere ions and the membrane forbidden. A cylindrical pore of length $H^{\mathrm{P}}$ spans the membrane and is bounded radially by a hard wall at $R^{\mathrm{P}}=1.5$ nm.
This value is a compromise; smaller values would have prevented the contact of Ca$^{2+}$ and Na$^{+}$ ions with some of the silanol oxygens (the wall of the MD pore model is ragged, see Fig.~\ref{fig1}). 
 
The oxygen atoms of the silanol groups are modeled explicitly as charged hard spheres with a radius of $0.135$ nm. 
Following the approach of Finnerty et al.~\cite{Finnerty}, the oxygen atoms are confined by harmonic potentials around reference positions taken from the MD model, thereby allowing some flexibility of the surface groups.~\cite{lakics_jcp_2026}

We consider two models that differ in the charges assigned to the oxygen atoms. In the first model, each oxygen atom carries a charge of $-e$, thus mimicking the MD model in which the magnitude of the oxygen charge is the same as that of a Cl$^{-}$ ion. In the second model, the oxygen charge is reduced to $-0.74e$. 
Results for the latter case together with the corresponding MD model are reported in the SM.
 
We study this model using a hybrid method in which the NP equation (Eq.~\ref{eq:NP}) is coupled to the Local Equilibrium Monte Carlo (LEMC) method to solve the transport problem self-consistently.~\cite{boda_jctc_2012} 
The LEMC method operates in the grand canonical ensemble, allowing the number of ions to fluctuate within predefined volume elements according to the local chemical potential, $\mu_{i}(z,r)$. The primary output of the LEMC simulation is the concentration profile, $c_{i}(z,r)$, although other quantities, such as the electrical potential and excess chemical potential profiles, can also be obtained.

Given the relation between $\mu_{i}(z,r)$ and $c_{i}(z,r)$ provided by LEMC, together with a prescribed diffusion coefficient profile, $D_{i}(z,r)$, the particle current density profile, $\mathbf{j}_{i}(z,r)$, can be computed from the NP equation. The chemical potential profile is then iteratively adjusted until the resulting current density satisfies the steady-state continuity equation,
$\nabla\cdot\mathbf{j}_{i}(z,r)=0$.

Note that the LEMC simulations are performed in three-dimensional geometry. To keep the computational cost tractable, however, we assume rotational symmetry. Accordingly, the results of the 3D LEMC simulations are averaged over the azimuthal angle to obtain profiles in the $(z,r)$ plane, which are then used in the NP equation. The $(z,r)$ plane is discretized into rectangular elements of size $\Delta z \times \Delta r$, with $\Delta z,\Delta r\approx 0.2$ nm.

\subsection{Fitting the radial diffusion coefficient profile} 

To fit the reduced model to the all-atom model, we first need to identify a quantity that can be directly compared between the two descriptions. 
The transport feature we aim to reproduce with the reduced model is the spatial distribution of the local current density between the near-wall region, associated with surface conductance, and the central region of the pore, associated with volume conductance. This distribution is characterized by the particle current density profile, $j_{i}(z,r)$, inside the pore.

Because the surface charge density is approximately homogeneous along the pore, the transport properties exhibit only weak variation along the $z$-axis. We therefore average the profiles over the pore length, $H^{\mathrm{P}}$, and focus on radially resolved quantities, such as the particle current density profile $j_{i}(r)$.
In the following, whenever radial profiles of physical quantities are shown, they refer to the pore region; therefore, the superscript P is omitted for simplicity; the exception is $D_{i}^{\mathrm{P}}(r)$.

The applied voltage, $U$, however, differs between the two simulation methods. To enable a direct comparison, we therefore use the radial electrical conductivity profile inside the pore,
\begin{equation}
	\kappa_{i}(r)=\frac{q_{i}j_{i}(r)}{E^{\mathrm{P}}},
	\label{eq:kappadef}
\end{equation}
where $E^{\mathrm{P}}$ is the $z$-component of the average electric field inside the pore.
Because $D_{i}^{\mathrm{P}}$ is a scalar, the conductivity tensor $\kappa_i$ is isotropic; here, we consider its axial $(z,z)$ component.
This quantity can be computed directly in NP+LEMC, but the corresponding electric field is not directly available from the MD simulations. Therefore, a simple procedure has been devised to estimate $E^{\mathrm{P}}$ for both methods (see Appendix~\ref{appendix}).

While in our previous work \cite{shabbir_jcp_2026} we analyzed the results on the basis of
\begin{equation}
	j_{i}(r)=v_{i}(r)c_{i}(r),
\end{equation}
here we use quantities normalized by the electric field, resulting in
\begin{equation}
	\kappa_{i}(r)=q_{i}u_{i}(r)c_{i}(r),
	\label{eq:kappauc}
\end{equation}
where the local electrical conductivity, $\kappa_{i}(r)$, is defined in Eq.~\ref{eq:kappadef}, and $u_{i}(r)=v_{i}(r)/E^{\mathrm{P}}$ is the ionic mobility. These quantities are local in the radial direction while being averaged over the pore length, $H^{\mathrm{P}}$.

While this relation is straightforward in MD, deriving Eq.~\ref{eq:kappauc} from the NP equation requires some care. The radial conductivity profile is obtained as
\begin{align}
	\kappa_{i}(r)&=\frac{q_{i}j_{i}(r)}{E^{\mathrm{P}}} = \frac{q_{i}}{E^{\mathrm{P}}} \left( \frac{1}{H^{\mathrm{P}}} \int\limits_{H^{\mathrm{P}}} j_{i}(z,r) dz \right) \nonumber \\	
	 & =  \frac{q_{i}}{E^{\mathrm{P}}} \left( \frac{1}{H^{\mathrm{P}}} \int\limits_{H^{\mathrm{P}}} \left[ -\frac{D_{i}^{\mathrm{P}}(r)}{kT} c_{i}(z,r)\frac{\partial \mu_{i}(z,r)}{\partial z} \right] dz \right) ,
\end{align}
where the integral is taken over the pore length, $H^{\mathrm{P}}$.
In the NP equation, the flux is expressed as the product of three functions: $D^{\mathrm{P}}_{i}(r)$, $c_{i}(z,r)$, and $\partial\mu_{i}(z,r)/\partial z$. By construction, the diffusion coefficient is independent of $z$ inside the pore. For the gradient of the electrochemical potential along the $z$ direction, we make the following assumptions.

\begin{figure}[t!]
	\begin{center}
		\includegraphics*[width=0.45\textwidth]{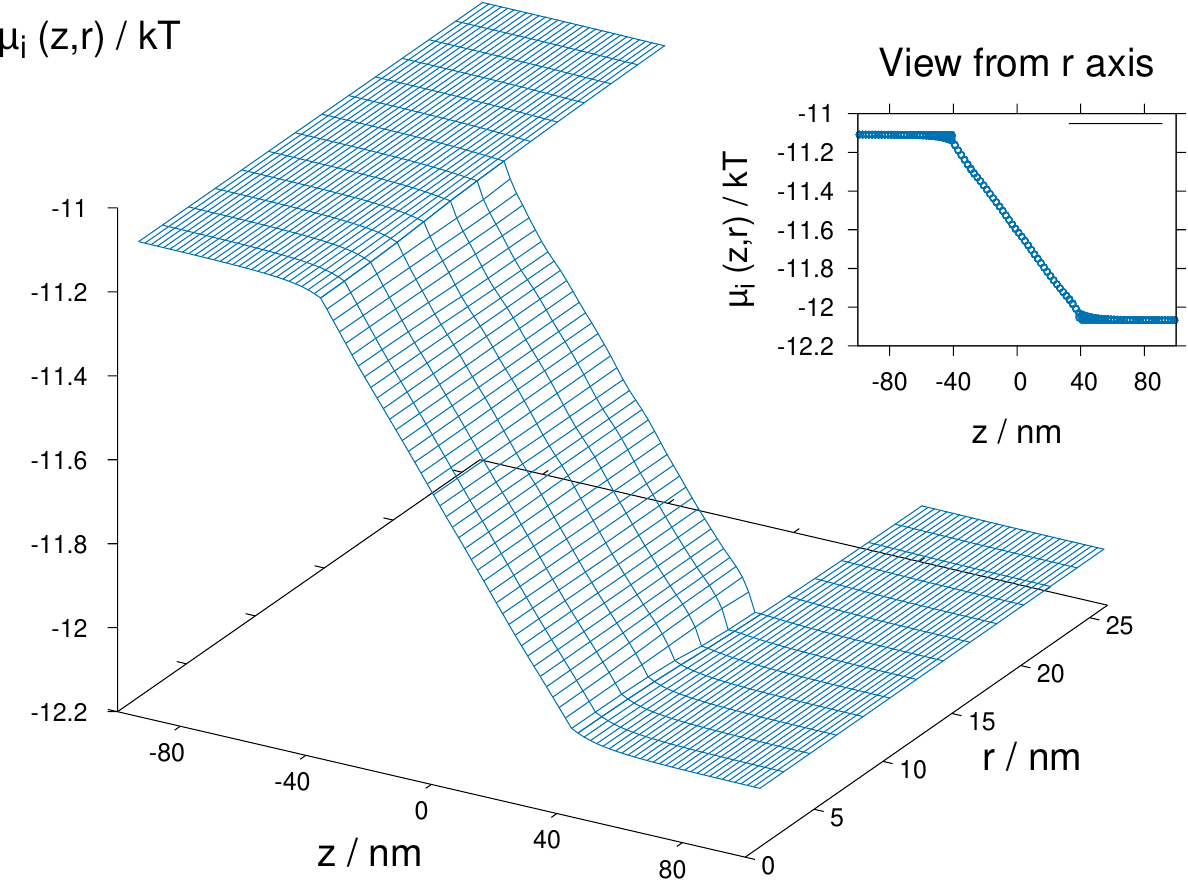}
	\end{center}
	\caption{Electrochemical potential over the $(z,r)$ plane (main panel) of Ca$^{2+}$ obtained from NP+LEMC and its view from the direction of the $r$ axis (subset).
	}
	\label{fig2}
\end{figure}

First, because the steady-state radial flux vanishes inside the pore, the NP equation implies $\partial\mu_{i}/\partial r=0$, and therefore we assume that $\mu_{i}$ is independent of $r$, namely, $\mu_i=\mu_i(z)$. Second, we assume that $\mu_{i}(z)$ decreases approximately linearly along the pore. Figure~\ref{fig2} shows that these are reasonable assumptions. As a result, we can write
\begin{align}
	\frac{\partial \mu_{i}(z)}{\partial z} & \approx \frac{ \left(kT\ln a_{i}^{\mathrm{R}}+q_{i}\phi^{R} \right)
		- \left( kT\ln a_{i}^{\mathrm{L}}+q_{i}\phi^{\mathrm{L}}\right)}{H^{\mathrm{P}} } \nonumber \\
	& = \frac{q_{i}U^{\mathrm{P}}}{H^{\mathrm{P}}} = -q_{i}E^{\mathrm{P}} ,
\end{align}
where the superscripts L and R denote the left and right ends of the pore, respectively, $U^{\mathrm{P}}=\phi^{\mathrm{R}}-\phi^{\mathrm{L}}$, and we have used the assumption that the activities are the same on the two sides of the membrane, $a_{i}^{\mathrm{L}}\approx a_{i}^{\mathrm{R}}$.

Combining these results, the radial conductivity profile becomes
\begin{align}
	\kappa_{i}(r) & =  \frac{q_{i}}{E^{\mathrm{P}}} \left( \frac{q_{i} E^{\mathrm{P}} D_{i}^{\mathrm{P}}(r)}{H^{\mathrm{P}}kT} \int\limits_{H^{\mathrm{P}}} c_{i}(z,r)dz \right)  \nonumber \\
	& = q_{i} \left( \frac{q_{i} D_{i}^{\mathrm{P}}(r)}{kT} \right) \frac{1}{H^{\mathrm{P}}} \int\limits_{H^{\mathrm{P}}} c_{i}(z,r)dz \nonumber \\
	& = q_{i} u_{i}(r)c_{i}(r),
\end{align}
where we have used the Nernst--Einstein relation, 
\begin{equation}
	u_{i}(r)=\frac{q_{i}D_{i}^{\mathrm{P}}(r)}{kT} ,
	\label{eq:13}
\end{equation}
for the mobility. 
At the same time, we can also compute mobility as $u_{i}(r)=\kappa_{i}(r)/q_{i}c_{i}(r)$, which gives the same result as the Nernst--Einstein relation provided that the approximation $\partial \mu_{i}/\partial z = -q_{i}E^{\mathrm{P}}$ holds. 
We verified this relation and found very good agreement (a representative case is shown in the SM), indicating that the above approximation for $\partial \mu_{i}/\partial z $ is satisfactory.

This result also underlines the approximate nature of the NP equation, which is formulated separately for each ionic species. In particular, direct momentum exchange between different ionic species is not explicitly included, representing a major approximation. 
Any coupling between ionic species arising from the high electrolyte concentration and charged confinement must therefore be incorporated into the effective diffusion coefficient, $D_{i}^{\mathrm{P}}(r)$.

\section{Results and Discussion} 

\begin{figure*}[t!]
	\begin{center}
		\includegraphics*[width=0.7\textwidth]{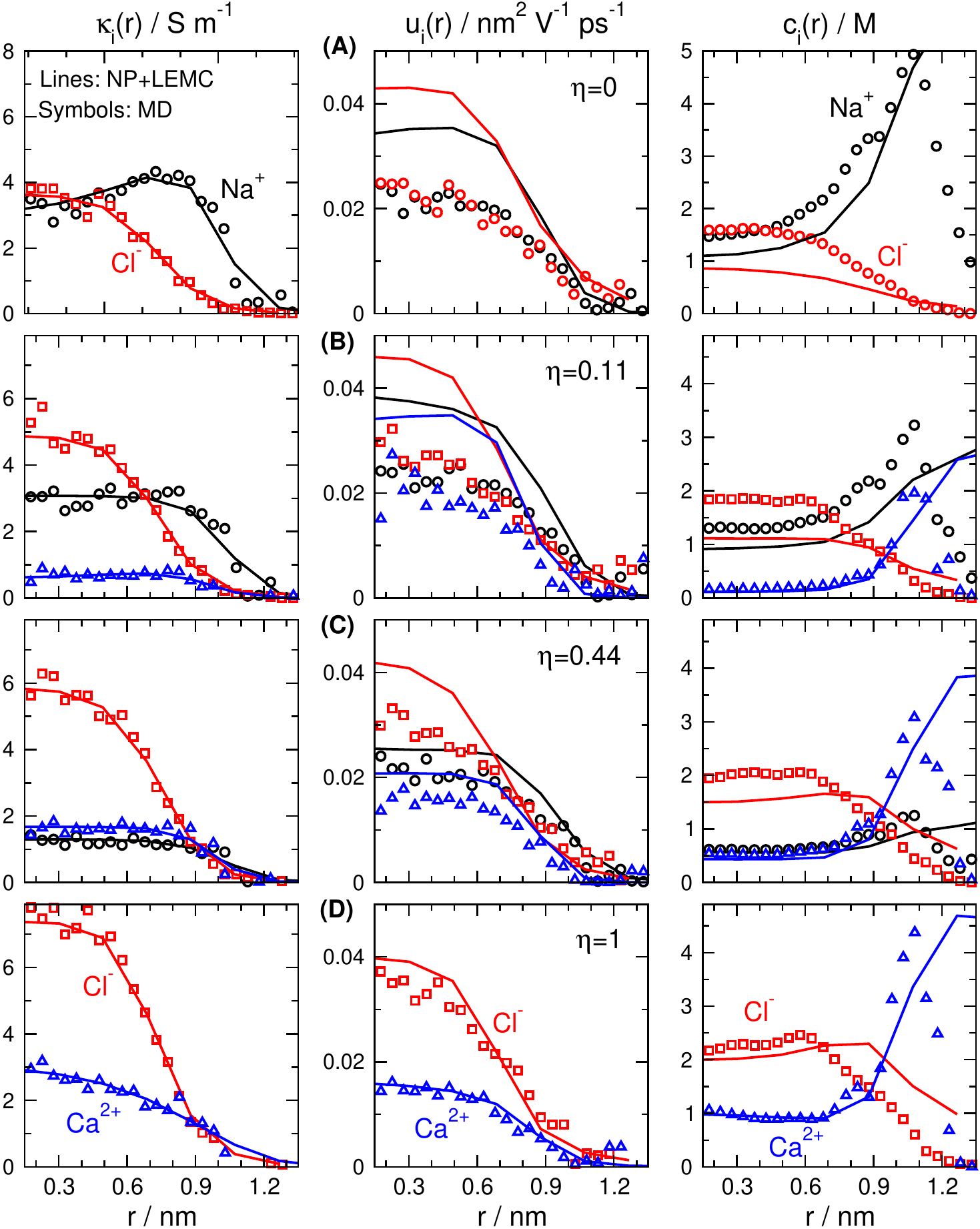}
	\end{center}
	\caption{From left to right: radial dependence of the axial components of electrical conductivity (obtained from the $z$-components of the flux density and electric field, see Eq.~\ref{eq:kappadef}; unit is S/m), mobility (in nm$^{2}$/V\,ps), and concentration (in mol/dm$^{3}$) for mixtures of NaCl and CaCl$_{2}$.
	Rows (A) to (D) refer to Ca$^{2+}$ mole fractions $\eta=0$, $0.11$, $0.44$, and $1$, respectively.
	Results for other mole fractions are found in the SM.
	Black, blue, and red colors refer to Na$^{+}$, Ca$^{2+}$, and Cl$^{-}$ ions, respectively.
	The figure shows results for the full-charge silanol oxygen (O$_{\mathrm{S}}$) model.
	Results for the scaled-charge O$_{\mathrm{S}}$ model are found in the SM.
	Symbols and lines refer to MD and NP+LEMC results, respectively. 
	}
	\label{fig3}
\end{figure*} 

During the fitting of the MD-derived $\kappa_{i}(r)$ profiles, we had to decide whether to fit the reduced model directly to the noisy profiles obtained from MD or to smooth functions fitted to the MD data. We chose the latter approach and fitted the asymmetric sigmoid function
\begin{equation}
	f(x)=\frac{\alpha_{1}+\alpha_{5}x^{2}}{\left[ \alpha_{2}+\exp (\alpha_{3}x) \right]^{\alpha_{4}}}
	\label{eq:sigmoid}
\end{equation}
which proved to provide an appropriate representation of the MD data (the $\alpha_{5}x^{2}$ term can produce a maximum in a few cases). 
In Fig.~\ref{fig3}, the $\kappa_{i}(r)$ profiles obtained from NP+LEMC (lines) therefore correspond to these fitted functions, since the latter were used as the target profiles in fitting the $D_{i}^{\mathrm{P}}(r)$ profiles.

In Fig.~\ref{fig3}, we show the radial profiles of $\kappa_{i}(r)$, $u_{i}(r)$, and $c_{i}(r)$ in the three columns for various mole fractions of NaCl and CaCl$_{2}$ in rows (A)--(D). 
In each case, the total salt concentration, $c_{\mathrm{tot}}=[\mathrm{NaCl}]+[\mathrm{CaCl}_{2}]$, 
is kept approximately constant at 1 M. 
The small deviations from the value of $c_{\mathrm{tot}}=1$ M arise from the use of canonical MD simulations, in which the ionic concentrations can only be adjusted approximately through a limited number of iterations.

In essence, the MD results for pure NaCl and CaCl$_{2}$ [rows (A) and (D)] have already been published in our previous paper~\cite{shabbir_jcp_2026}, but were presented there in terms of the $j_{i}(r)$ and $v_{i}(r)$ profiles. 
Here, we present the same results in terms of the $\kappa_{i}(r)$ and $u_{i}(r)$ profiles, together with the corresponding NP+LEMC results. The latter were obtained by fitting the $D_{i}^{\mathrm{P}}(r)$ profiles to the smoothed MD $\kappa_{i}(r)$ profiles using Eq.~\ref{eq:sigmoid}.

As in our previous work,~\cite{shabbir_jcp_2026} a clear distinction can be made between the transport behavior of NaCl and CaCl$_{2}$. 
NaCl exhibits classical EDL behavior, with an excess of relatively mobile Na$^{+}$ ions near the pore wall. 
Although Na$^{+}$ ions bind to the silanol groups, their mobility remains comparable to that of Cl$^{-}$ ions. 
Consequently, the Na$^{+}$ contribution to the electrical conductance is larger than that of Cl$^{-}$ inside the pore. 
This excess Na$^{+}$ conductance originates from the EDL region, while the central region of the pore exhibits similar conductivities for Na$^{+}$ and Cl$^{-}$.

The negatively charged silica nanopore is therefore cation selective for NaCl in both the equilibrium and dynamical senses. 
In equilibrium, more Na$^{+}$ ions than Cl$^{-}$ ions accumulate in the near-wall region, whereas dynamically, Na$^{+}$ ions carry a larger fraction of the electrical current.

CaCl$_{2}$ exhibits markedly different behavior because Ca$^{2+}$ ions are strongly bound to the silanol groups. 
As discussed in our previous work,~\cite{shabbir_jcp_2026} the following picture emerges for pure CaCl$_{2}$. 
Ca$^{2+}$ ions are highly concentrated in the surface layer, but their mobility is strongly suppressed there. 
Cl$^{-}$ ions have a much lower concentration in the surface layer and a somewhat higher, although still limited, mobility. 
As a result, both ionic species exhibit low conductivity in the surface layer, which therefore contributes little to the total conductance.

In contrast, an electrolyte with nearly bulk-like stoichiometry forms in the central region of the pore. 
The mobility of Cl$^{-}$ is much larger than that of Ca$^{2+}$, and consequently, the conductivity of Cl$^{-}$ is also larger than that of Ca$^{2+}$. 
This leads to a weak Cl$^{-}$ selectivity of the pore, in agreement with the experiments of He et al.~\cite{he_jacs_2009}.

\begin{figure}[t!]
	\begin{center}
		\includegraphics*[width=0.35\textwidth]{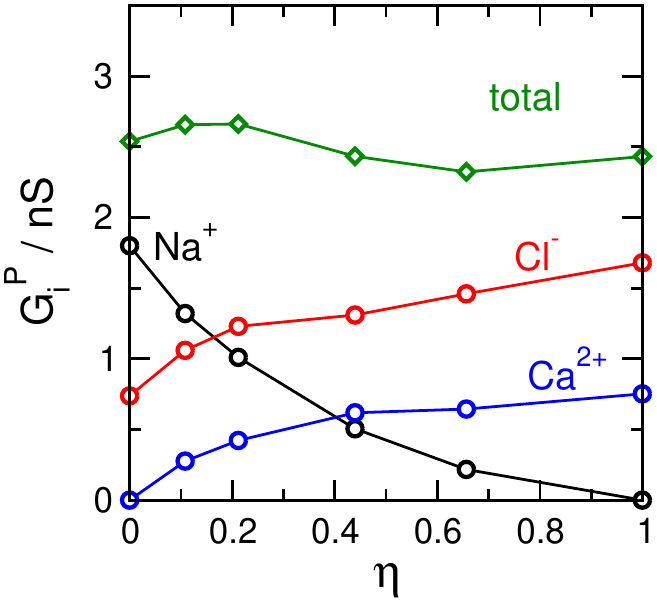}
	\end{center}
	\caption{Conductances of different ionic species and the total conductance (defined in Appendix~\ref{appendix}) as obtained from MD simulations as functions of Ca$^{2+}$ mole fraction.
	Black, blue, and red colors refer to Na$^{+}$, Ca$^{2+}$, and Cl$^{-}$ ions, respectively.
	Error bars are smaller than the size of the symbols.
	}
	\label{fig4}
\end{figure} 

After discussing the pure NaCl and CaCl$_{2}$ electrolytes, we continue with studying the composition dependence.
For the mixtures [rows (B) and (C)], the profiles change systematically with composition: as the CaCl$_{2}$ mole fraction, $\eta$, increases, both $c_{i}(r)$ and $\kappa_{i}(r)$ increase for Ca$^{2+}$, while they decrease for Na$^{+}$. 

This dependence is not strictly linear, as can also be seen in Fig.~\ref{fig4}, which shows the ionic and total conductances (defined in Appendix~\ref{appendix}) as functions of $\eta$. 
The nonlinearity is apparent and can be attributed to the preferential binding of Ca$^{2+}$ over Na$^{+}$ to the silanol groups.

As Ca$^{2+}$ ions are introduced into the mixture, they disproportionately replace Na$^{+}$ ions inside the pore, as illustrated by the concentration profiles in Fig.~\ref{fig3}B. 
Even though Ca$^{2+}$ accounts for only 11\% of the salt concentration in this mixture, its concentration peak near the wall (blue triangles) is already nearly as high as that of Na$^{+}$ (black circles). 
Since the mobilities are less strongly affected by the composition, the Ca$^{2+}$ and Na$^{+}$ conductivities follow the corresponding concentration trends.
Although this composition dependence is more pronounced in the EDL layer, its contribution to the total current is relatively small, so its effect on the overall nonlinearity in Fig.~\ref{fig4} is also small.
Consequently, only a weak nonlinearity is observed in the $G_{i}(\eta)$ curves.
This nonlinearity is even weaker in the scaled-charge O$_{\mathrm{S}}$ model (see Fig.~S8 of the SM).

At the same time, Cl$^{-}$ ions gradually become the dominant charge carriers as the Ca$^{2+}$ mole fraction increases. 
In the pure CaCl$_{2}$ system, Cl$^{-}$ ions carry approximately twice as much current as Ca$^{2+}$ ions.

The combination of these opposing trends among the three ionic species results in a weak maximum in the total conductance.
This behavior is qualitatively different from the classical AMFE, which is characterized by a minimum in the $G(\eta)$ curve.
Such a minimum is practically absent in the present high-concentration system, indicating that AMFE does not occur under these conditions.
In contrast, AMFE has been observed at substantially lower concentrations, typically $\leq 0.1$ M.~\cite{almers_jp_1984a,almers_jp_1984b,gillespie_jpcb_2005,gillespie_bj_2008_energetics,gillespie_giri_bj_2009,gillespie_bj_2008_nanopore}

\begin{figure}[t!]
	\begin{center}
		\includegraphics*[width=0.3\textwidth]{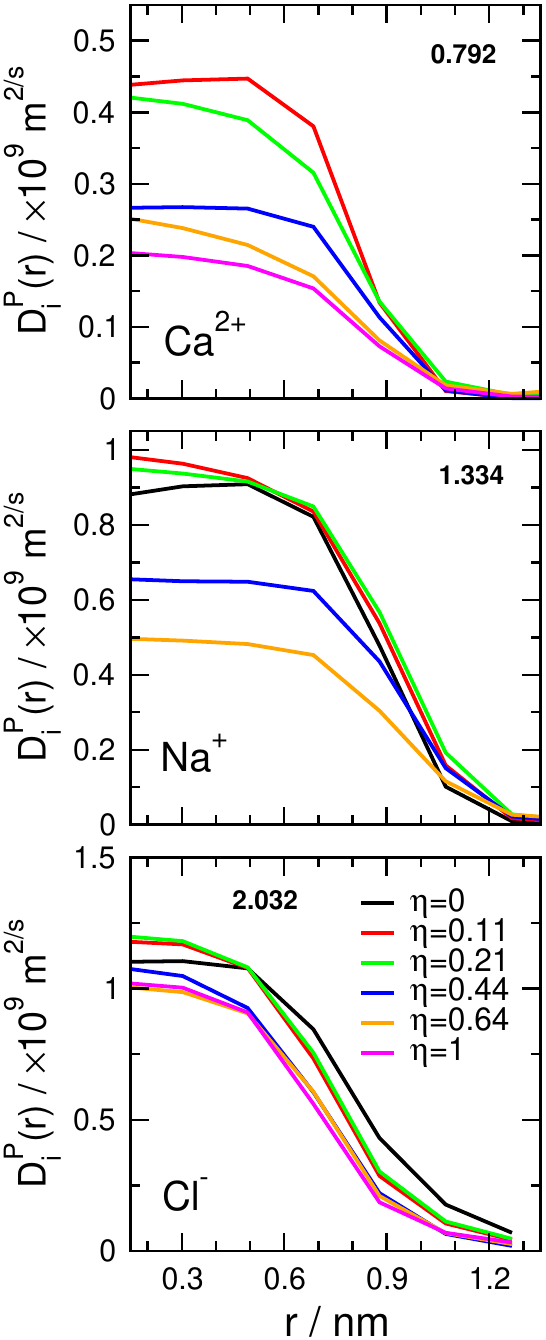}
	\end{center}
	\caption{Radial diffusion coefficient profiles in the pore as obtained from fitting the NP+LEMC conductivity profiles to  MD conductivity profiles.
	Different panels refer to different ionic species, while different colors refer to different mole fractions.
	The numbers in the panels indicate the bulk values, $D_{i}^{\mathrm{B}}$.
	}
	\label{fig5}
\end{figure} 

In Fig.~\ref{fig3}, the conductivity profiles obtained from NP+LEMC agree well with the MD profiles, as expected, since the fitting procedure has been designed to reproduce them. 
The mobility profiles, however, do not agree quantitatively because the concentration profiles also differ between the two models (see Eq.~\ref{eq:kappauc}). 
The fitted diffusion coefficient profiles must therefore account for the differences in the microscopic descriptions of the two models, which give rise to different ionic distributions. 
Nevertheless, the qualitative behavior is the same in both methods: the mobility decreases toward the pore wall.

The values of the fitted diffusion coefficient profiles in the confined nanopore (Fig.~\ref{fig5}) are systematically smaller than the corresponding bulk values, $D_{i}^{\mathrm{P}}(r)<D_{i}^{\mathrm{B}}$.
The overall shape of the profiles is similar for all ionic species and mole fractions. 
The main difference lies in the value of $D_{i}^{\mathrm{P}}(r)$ in the central region of the pore, $r\rightarrow0$. 
An interesting result is that the fitted $D_{i}^{\mathrm{P}}(r)$ profiles decrease for all ionic species as $\eta$ increases.

To understand this behavior, we return to Fig.~\ref{fig3}. The mobility profiles obtained from MD and NP+LEMC differ more strongly at low values of $\eta$. This trend follows the behavior of the concentration profiles: interestingly, the MD concentration profiles are reproduced more accurately by the reduced model at large Ca$^{2+}$ concentrations. This may be related to the increasing ionic strength upon addition of divalent Ca$^{2+}$ ions, which leads to a shorter Debye screening length and thus more efficient screening of the charged nanopore by the electrolyte inside the pore. Consequently, less screening by ions from the surrounding bulk solution is required, and the resulting ionic distribution inside the nanopore is closer to charge neutrality and more consistent with the MD results.

In summary, the limited transferability of the $D_{i}^{\mathrm{P}}(r)$ profiles across different compositions stems from the fact that the reduced model reproduces the MD concentration profiles with different degrees of accuracy at different compositions.

\section{Summary}

We have developed a multiscale approach for incorporating molecular-scale transport information into a computationally efficient reduced model of ion transport in nanopores. 
The approach uses explicit-water MD simulations to obtain radially resolved conductivity profiles, $\kappa_i(r)$, which are then used as target functions for fitting radially varying effective diffusion coefficients, $D_i^{\mathrm{P}}(r)$, in an implicit-water NP+LEMC model. 
In contrast to fitting a single, spatially averaged transport coefficient, this procedure preserves the radial structure of ionic transport and allows the reduced model to reproduce the local current distribution observed in the higher-resolution model.

\begin{figure*}[t!]
	\begin{center}
		\includegraphics*[width=0.75\textwidth]{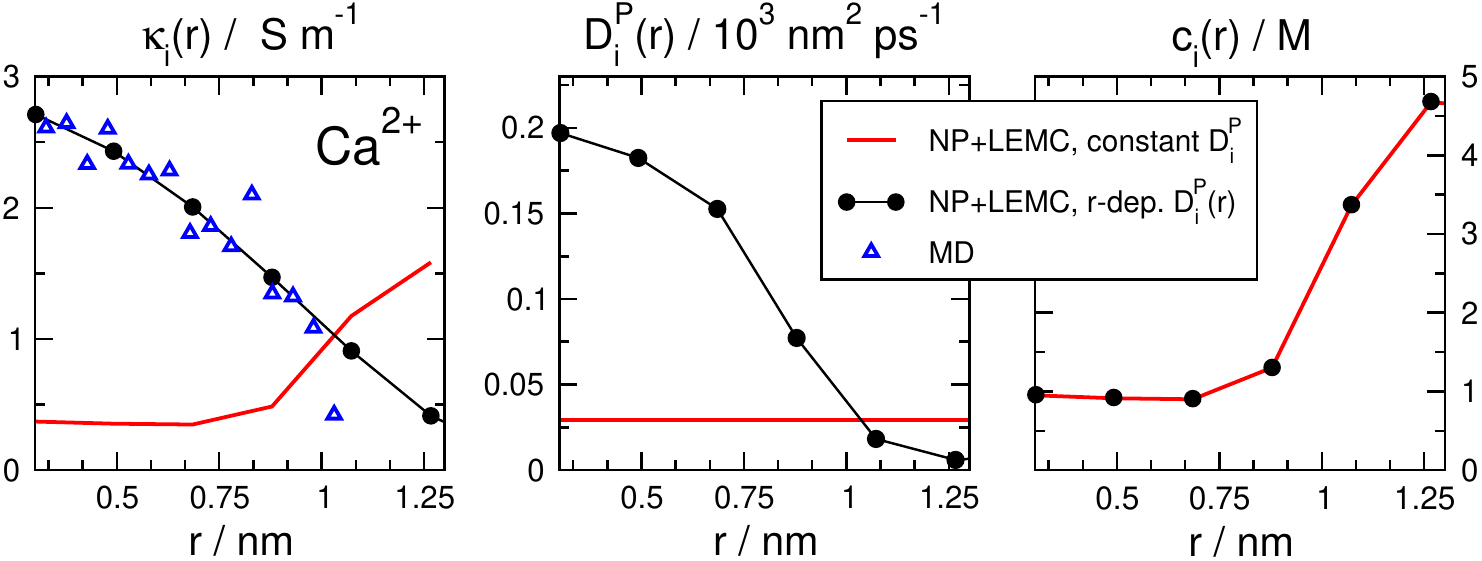}
	\end{center}
	\caption{From left to right: radial dependence of the electrical conductivity, diffusion coefficient, and concentration inside the pore for Ca$^{2+}$ in the $\eta=1$ case.
	Red curves refer to the case in which the constant value of average diffusion coefficient inside the pore is fitted. 
	Black curves with circles refer to the case in which the radially varying $D_{i}^{\mathrm{P}}(r)$ profile is fitted to the MD $\kappa_{i}(r)$ profile (blue triangles). 
	}
	\label{fig6}
\end{figure*}

The importance of this distinction is illustrated in Fig.~\ref{fig6}. 
When only a constant, average diffusion coefficient is fitted (red color), the conductivity profile follows the concentration profile and therefore increases toward the pore wall. 
This behavior is qualitatively different from that observed in the explicit-water MD simulations (blue triangles), where the conductivity decreases near the wall because of strongly suppressed ionic mobility. 
Fitting a radially varying $D_i^{\mathrm{P}}(r)$ reverses this behavior and reproduces the MD conductivity profile, while leaving the concentration profile essentially unchanged (black circles). 
Thus, the radial diffusion coefficient is not merely a fitting device for the overall conductance; it is required to reproduce the spatial distribution of local transport.

Application to NaCl, CaCl$_2$, and their mixtures in a negatively charged silica nanopore demonstrates the physical consequences of this approach. 
NaCl exhibits classical EDL behavior, with an excess of relatively mobile Na$^+$ ions near the pore wall and enhanced cation conduction. 
In CaCl$_2$, strong binding of Ca$^{2+}$ to the silanol groups produces a highly concentrated but nearly immobile surface layer, so that this region contributes little to the total conductance. 
The ionic conductances in the mixtures vary systematically but nonlinearly with composition, because Ca$^{2+}$ preferentially binds to the silanol groups and therefore replaces Na$^+$ in the pore more strongly than expected from the bulk composition.
Across all compositions, the fitted diffusion coefficients decrease toward the pore wall, reflecting the suppressed mobility observed in the explicit-water simulations.

A legitimate criticism of reduced device models is their black-box nature: 
although the macroscopic or device-level behavior can often be reproduced efficiently by adjusting effective parameters, the physical meaning of these parameters is not always clear.
An effective diffusion coefficient, for example, may simultaneously account for confinement, solvent effects, surface interactions, polarization, and other molecular-scale processes that are treated explicitly only at higher resolution. 
While this can be advantageous for reproducing device-level observables, it can also obscure the underlying molecular mechanisms and limit the predictive value and transferability of the reduced model.

The present approach addresses this limitation by introducing a higher-resolution description between the reduced model and experiment, thus creating a bridge between molecular-level mechanisms and device-level behavior. 
Rather than fitting the effective parameters of the reduced model directly to integrated experimental observables, the higher-resolution model provides spatially resolved target functions from which these parameters can be determined. 

In the present case, explicit-water MD simulations provide the radial conductivity profiles that serve as targets for fitting the effective radial diffusion coefficient of the implicit-water NP model. 
The higher-resolution model thus serves as an intermediate layer between experimentally relevant device behavior and the computationally efficient reduced description, allowing molecular-scale transport information to be incorporated into the effective diffusion coefficients.

In the present work, we demonstrate the basic idea using MD-derived radial conductivity profiles as target quantities; the experimental level is not yet included.
In the continuation of this work, we will combine both levels of information by fitting the diffusion coefficient profiles to experimental and MD data simultaneously.
Experimental conductance data will provide information about the integrated, device-level transport, while radial current-density profiles from MD will provide information about the spatially resolved, molecular-scale transport within the pore.
Together, these two fitting procedures will determine the magnitude of the transport from experimental conductance data and its spatial profile (``shape'') from the radial MD data.

\section*{Supplementary Material}
\label{si}
The Supplementary Material contains results ($\kappa_{i}(r)$, $u_{i}(r)$, $c_{i}(r)$, $D_{i}^{\mathrm{P}}(r)$, and $G_{i}(\eta)$) for the scaled-charge O$_{\mathrm{S}}$ model, as well as profiles for the full-charge O$_{\mathrm{S}}$ model that were not included in Fig.~3 of the paper.
Comparison of the mobility curves computed from Eq.~\ref{eq:kappauc} and Eq.~\ref{eq:13} is shown.

\section*{Acknowledgements}

We gratefully acknowledge  the financial support of the National Research, Development and Innovation Office -- NKFIH K124353.
We acknowledge KIFÜ (Governmental Agency for IT Development, Hungary, \url{https://ror.org/01s0v4q65}) for awarding us access to the Komondor HPC facility based in Hungary.
The authors are grateful to Dávid Fertig for helpful discussions.

\appendix
\section{Derivation of average electric field strength in the pore}
\label{appendix}

\begin{figure}[t!]
	\begin{center}
		\includegraphics*[width=0.4\textwidth]{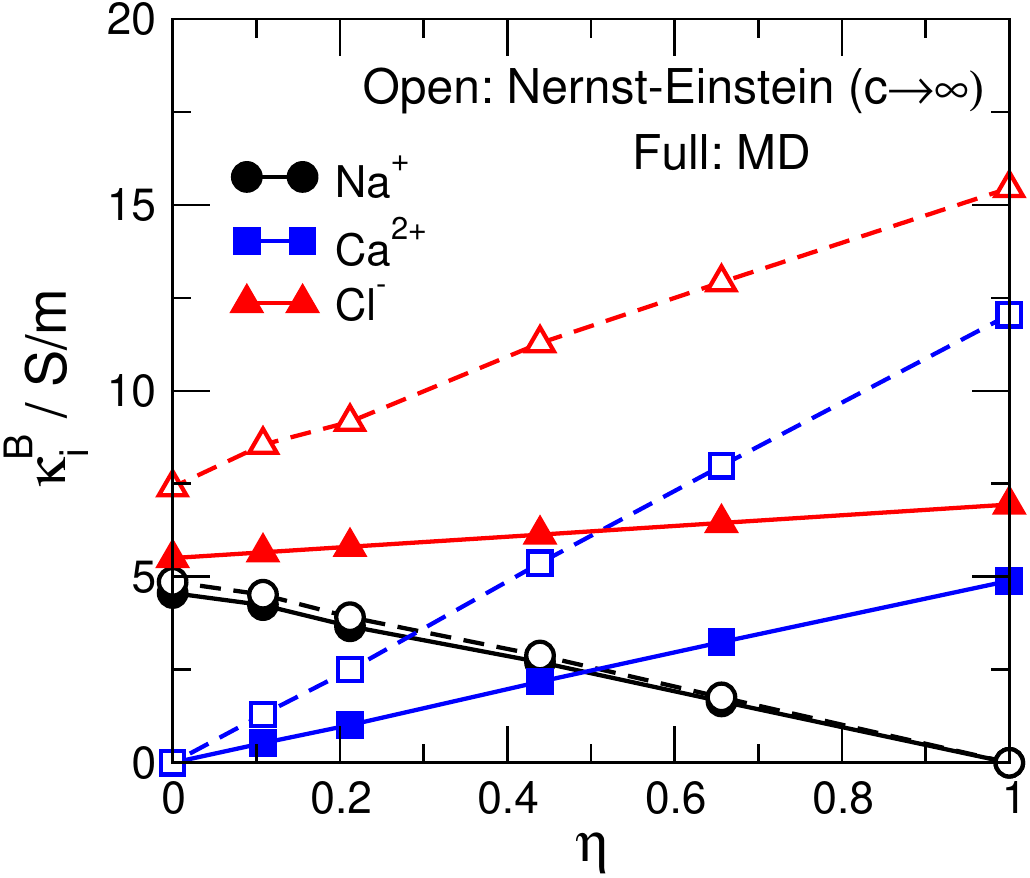}
	\end{center}
	\caption{Ionic and total bulk conductivities as functions of Ca$^{2+}$ mole fractions for the various mixtures. 
		Full symbols represent results obtained from bulk MD simulations for the given composition. 
		Open symbols are values calculated from the Nernst-Einstein relation (Eq.~\ref{eqapp:nernst-einstein}) using experimental self diffusion constant data, $D_{i}^{\mathrm{B}}$. Different colors denote different ionic species.
		The error bars are on the order of the symbol size.
	}
	\label{fig7}
\end{figure} 

We estimate the average electric field strength in the pore, $E^{\mathrm{P}}$, using the following procedure.
The simulation cell is divided axially into pore (P) and bulk (B) regions with lengths $H^{\mathrm{P}}$ and $H^{\mathrm{B}}$ and cross-sectional areas $A^{\mathrm{P}}$ and $A^{\mathrm{B}}$, respectively.
Note that the cross-sectional area is that of a square in the MD cell ($A^{\mathrm{B}}=L^{2}$), while it is that of a circle in the LEMC cell ($A^{\mathrm{B}}=\pi(R^{\mathrm{B}})^{2}$).

Under steady-state conditions, the electrical current $I_{i}=q_{i}J_{i}$ is the same in the two regions.
Therefore, ionic conductances in the pore and in the bulk are defined as  $G_{i}^{\mathrm{P}}=I_{i}/U^{\mathrm{P}}$ and $G_{i}^{\mathrm{B}}=I_{i}/U^{\mathrm{B}}$, respectively.
The total pore and bulk conductances are defined as $G^{\mathrm{P}}=I/U^{\mathrm{P}}$ and $G^{\mathrm{B}}=I/U^{\mathrm{B}}$, respectively, where $I=\sum_{i}I_{i}$ is the total current.
The voltage drop across the pore can then be estimated as
\begin{equation}
	U^{\mathrm{P}} = U-U^{\mathrm{B}} = U - \frac{I}{G^{\mathrm{B}}} = U-\dfrac{I}{\kappa^{\mathrm{B}}A^{\mathrm{B}}/H^{\mathrm{B}}} ,
\end{equation}
which leaves the conductivity in the bulk, $\kappa^{\mathrm{B}}$, to be estimated.

For the all-atom model, the bulk conductivity, $\kappa^{\mathrm{B}}$, is obtained from separate bulk MD simulations.
For the reduced model, it is calculated using the Nernst--Einstein relation,
\begin{equation}
	\kappa^{\mathrm{B}} = \sum_{i}\kappa^{\mathrm{B}}_{i} = \sum_{i} c_{i}^{\mathrm{B}} \frac{q_{i}^{2}D_{i}^{\mathrm{B}}}{kT} ,
	\label{eqapp:nernst-einstein}
\end{equation}
where $c_{i}^{\mathrm{B}}$ is the bulk concentration of ionic species $i$ at a given composition, $\eta$. 
Finally, the magnitude of the electric field inside the pore is estimated as $E^{\mathrm{P}}=U^{\mathrm{P}}/H^{\mathrm{P}}$.

The bulk conductivities obtained from bulk MD simulations and from the Nernst--Einstein relation are compared in Fig.~\ref{fig7}.
The deviations between the two estimates increase with increasing Ca$^{2+}$ concentration, reflecting the stronger ionic correlations induced by the presence of Ca$^{2+}$ ions, which suppress the bulk conductivity relative to the Nernst--Einstein prediction obtained by using the infinite-dilution diffusion coefficient, $D_i^{\mathrm{B}}$, at the bulk concentration $c_i^{\mathrm{B}}$.

A comparison of the pore conductivity profiles, $\kappa_i(r)$, shown in Fig.~\ref{fig3}, with the corresponding bulk conductivities, $\kappa_i^{\mathrm{B}}$, shown in Fig.~\ref{fig7}, shows that the ionic conductivities inside the pore are generally suppressed relative to their bulk values, primarily due to confinement and interactions with the pore charges.

Note that the electrical potential drop across the bulk, $U^{\mathrm{B}}$, is ultimately much smaller than that across the pore, $U^{\mathrm{P}}$ (a few percent of it), because the pore conductance is considerably smaller than the bulk conductance ($G^{\mathrm{P}}\ll G^{\mathrm{B}}$).
Therefore, using the total voltage $U$ instead of $U^{\mathrm{P}}$ would provide a good approximation.
Although this choice would lead to slightly different numerical results, it would not affect any of our qualitative conclusions.

\bibliography{nanopore,book,own}
\bibliographystyle{unsrt} 

\clearpage
\foreach \x in {1,...,6}{%
	\clearpage
	\includepdf[pages={\x}]{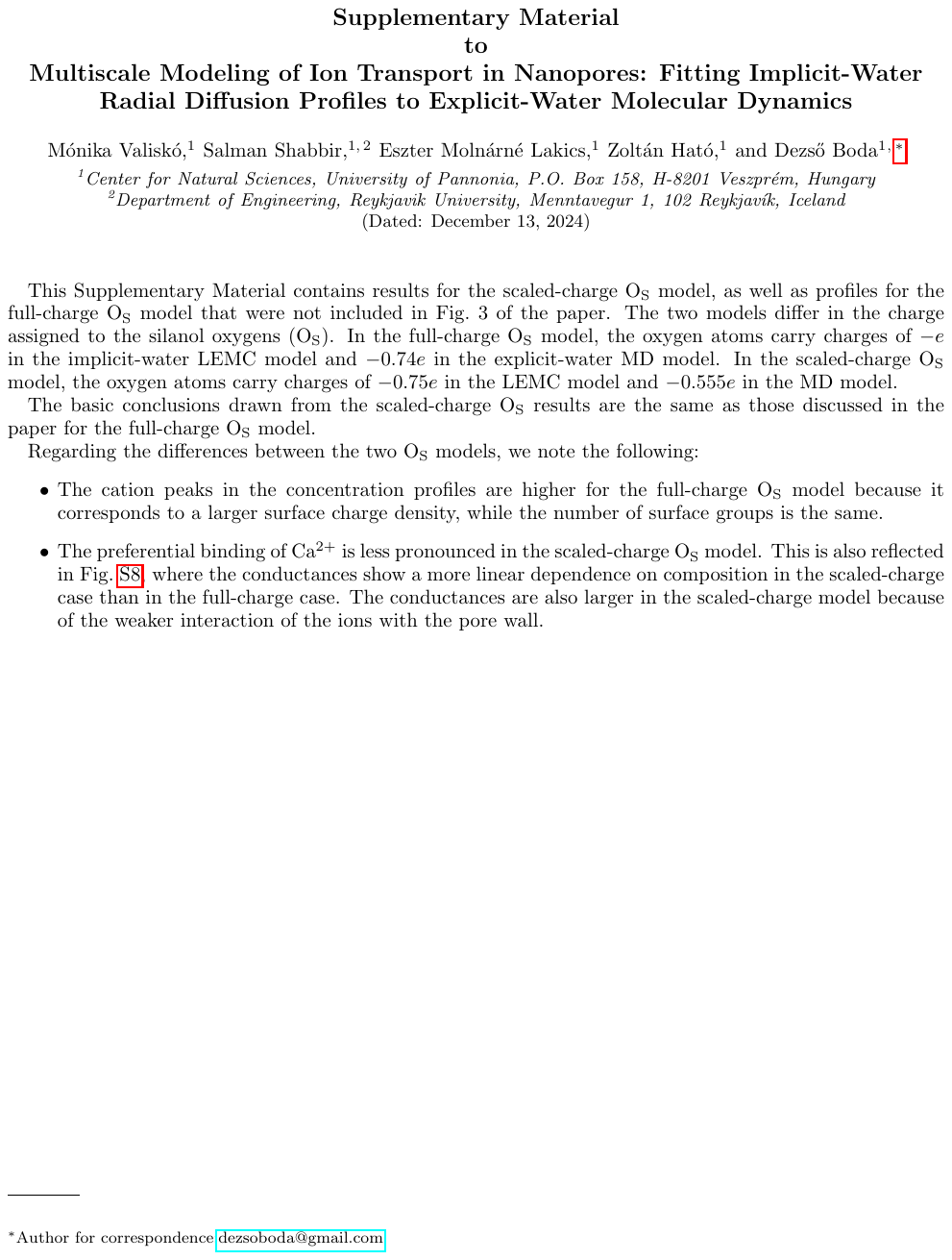}%
}
%\includepdf[pages=-]{Salman3-SI.pdf}
\end{document}